\documentclass{llncs}

\usepackage{ifthen,float}
\usepackage{amsmath}
\usepackage{color,xcolor}
\usepackage{hyperref} \hypersetup{colorlinks=true, breaklinks=true, urlcolor=black, linkcolor=black, citecolor=black}
\usepackage{amssymb}
\usepackage{subcaption}

\usepackage{tikz}
\usetikzlibrary{patterns}
\pgfdeclarepatternformonly{stripes}{\pgfqpoint{-4pt}{-4pt}}{\pgfqpoint{8pt}{8pt}}{\pgfqpoint{4pt}{4pt}}%
{
  \pgfsetlinewidth{1.414pt}
  \pgfpathmoveto{\pgfqpoint{0pt}{0pt}}
  \pgfpathlineto{\pgfqpoint{4pt}{4pt}}
  \pgfusepath{stroke}
}

\colorlet{colorA}{black!20!red}
\colorlet{colorB}{black!20!blue}

\usepackage[boxruled,vlined,linesnumbered]{algorithm2e}
\SetKwBlock{SubAlgoBlock}{}{end}
\newcommand{\SubAlgo}[2]{#1 \SubAlgoBlock{#2}}
\let\oldnl\nl
\newcommand{\nonl}{\renewcommand{\nl}{\let\nl\oldnl}}

\SetKw{Var}{var}
\SetKw{Op}{operation}
\SetKw{RecA}{when a message}
\SetKw{RecB}{is received from}
\SetKw{Broadcast}{FIFO broadcast}
\SetKw{Wait}{wait until}

\newcommand{\DataFont}[1]{\textsf{#1}}
\newcommand{\FunctionFont}[1]{\texttt{#1}}
    
\newcommand{\VarNameX}{X}          \SetKwData{KwX}{\VarNameX}   \newcommand{\AlgoX}{\DataFont{\VarNameX}}
\newcommand{\VarNameVC}{ValClock}  \SetKwData{KwVC}{\VarNameVC} \newcommand{\AlgoVC}{\DataFont{\VarNameVC}}
\newcommand{\VarNameSC}{SendClock} \SetKwData{KwSC}{\VarNameSC} \newcommand{\AlgoSC}{\DataFont{\VarNameSC}}
\newcommand{\VarNameG}{G}          \SetKwData{KwG}{\VarNameG}   \newcommand{\AlgoG}{\DataFont{\VarNameG}}
\newcommand{\VarNameV}{V}          \SetKwData{KwV}{\VarNameV}   \newcommand{\AlgoV}{\DataFont{\VarNameV}}

\newcommand{\VarNamev}{v}   \SetKwData{Kwv}{\VarNamev}   \SetKwFunction{KwGV}{\VarNamev} \newcommand{\Algov}{\DataFont{\VarNamev}} \newcommand{\AlgoGV}{\FunctionFont{\VarNamev}}
\newcommand{\VarNamek}{k}   \SetKwData{Kwk}{\VarNamek}   \SetKwFunction{KwGK}{\VarNamek} \newcommand{\Algok}{\DataFont{\VarNamek}} \newcommand{\AlgoGK}{\FunctionFont{\VarNamek}}
\newcommand{\VarNamet}{t}   \SetKwData{Kwt}{\VarNamet}   \SetKwFunction{KwGT}{\VarNamet} \newcommand{\Algot}{\DataFont{\VarNamet}} \newcommand{\AlgoGT}{\FunctionFont{\VarNamet}}
\newcommand{\VarNamecl}{cl} \SetKwData{Kwcl}{\VarNamecl} \SetKwFunction{KwGCL}{\VarNamecl} \newcommand{\Algocl}{\DataFont{\VarNamecl}} \newcommand{\AlgoGCL}{\FunctionFont{\VarNamecl}}

\newcommand{\VarNameWrite}{update}   \SetKwFunction{KwWrite}{\VarNameWrite} \newcommand{\AlgoWrite}{\FunctionFont{\VarNameWrite}}
\newcommand{\VarNameSnap}{snapshot} \SetKwFunction{KwSnap}{\VarNameSnap}  \newcommand{\AlgoSnap}{\FunctionFont{\VarNameSnap}}
\newcommand{\VarNameM}{message}     \SetKwFunction{KwM}{\VarNameM}     \newcommand{\AlgoM}{\FunctionFont{\VarNameM}}
    
\newcommand{\Algog}{g}
\newcommand{\REG}{\DataFont{REG}}

\makeatletter
\renewcommand{\algocf@caption@boxruled}{%
  \hrule
  \hbox to \hsize{%
    \vrule\hskip-0.4pt
    \vbox{
       \vskip\interspacetitleboxruled%
       \unhbox\algocf@capbox\hfill
       \vskip\interspacetitleboxruled
       }%
     \hskip-0.4pt\vrule%
   }\nointerlineskip%
}%
\makeatother

\begin{document}

\title{On Composition and Implementation \\ of Sequential Consistency}
\author{Matthieu Perrin \and Matoula Petrolia \and Achour Most\'efaoui \and Claude Jard}
\institute{LINA -- University of Nantes, France \hspace{7mm}\email{first.last@univ-nantes.fr}}

\maketitle
\begin{abstract}
To implement a linearizable shared memory in synchronous message-passing systems it is necessary to wait for a time linear to the uncertainty in the latency of the network for both read and write operations. Waiting only for one of them suffices for sequential consistency. 

This paper extends this result to crash-prone asynchronous systems, proposing a distributed algorithm that builds a sequentially consistent shared snapshot  memory on top of an asynchronous message-passing system where less than half of the processes may crash. We prove that waiting is needed only when a process invokes a read/snapshot right after a write. 

We also show that sequential consistency is composable in some cases commonly encountered: 1) objects that would be linearizable if they were implemented on top of a linearizable memory become sequentially consistent when implemented on top of a sequential memory while remaining composable and 2) in round-based algorithms, where each object is only accessed within one round.

\end{abstract}

\sloppy

\section{Introduction}

A distributed system is abstracted as a set of entities (nodes, processes, agents, etc) that communicate through a communication medium. 
The two most used communication media are communication channels (message-passing system) and shared memory (read/write operations). 
Programming  with shared objects is generally more convenient as it offers a higher level of abstraction to the programmer,
therefore facilitates the work of designing distributed applications. 
A natural question is the level of consistency ensured by shared objects. 
An intuitive property is that shared objects should behave as if all processes accessed the same physical copy of the object. 
\emph{Sequential consistency}~\cite{lamport1979make} ensures that all the operations in a distributed history 
appear as if they were executed sequentially, in an order that respects the sequential order of each process (\emph{process order}). 

Unfortunately, sequential consistency is not composable: if a program uses two or more objects,
despite each object being sequentially consistent individually, the set of all objects may not be sequentially consistent.
\emph{Linearizability}~\cite{herlihy1990linearizability} overcomes this limitation by adding constraints on real time: each operation appears at a single 
point in time, between its start event and its end event. As a consequence, linearizability enjoys the locality property \cite{herlihy1990linearizability} that ensures its composability. 
Because of this composability, much more effort has been focused on linearizability than on sequential consistency so far. 
However, one of our contributions implies that in asynchronous systems where no global clock can be implemented to measure real time, a process cannot distinguish between a linearizable and a sequentially consistent execution, thus 
the connection to real time
seems to be a worthless --- though costly --- guarantee.

In this paper we focus on message-passing distributed systems. In such systems a shared memory is not a physical object; it has to be built using the underlying message-passing communication network. 
Several bounds have been found on the cost of sequential consistency and linearizability in synchronous distributed systems, where the transit time for any message is in a range $[d-u,d]$,
where $d$ and $u$ are called respectively the \emph{latency} and the \emph{uncertainty} of the network. Let us consider an implementation of a shared memory, and let $r$ (resp. $w$) be the 
worst case latency of any read (resp. write) operation. Lipton and Sandberg proved in \cite{lipton1988pram} that, if the algorithm implements a sequentially consistent memory, the inequality $r+w\geq d$ must hold.
Attiya and Welch refined this result in~\cite{attiya1994sequential}, proving that each kind of operations could have a 0-latency implementation for sequential consistency (though not both in the same implementation)
but that the time duration of both kinds of operations has to be at least linear in $u$ in order to ensure linearizability.

Therefore the following questions arise. Are there applications for which the lack of composability of sequential consistency is not a problem? For these applications, can we expect the same benefits in weaker message-passing models, such as asynchronous failure-prone systems, from using sequentially consistent objects rather than linearizable objects?

To illustrate the contributions of the paper, we also address a higher level operation: a snapshot operation \cite{Afek93} that allows to read in a single operation a whole set of registers. A sequentially consistent snapshot is such that the set of values it returns may be returned by a sequential execution. This operation is very useful as it has been proved \cite{Afek93} that linearizable snapshots can be wait-free implemented from single-writer/multi-reader registers. Indeed, assuming a snapshot operation does not bring any additional power with respect to shared registers. Of course this induces an additional cost: the best known simulation needs $O(n\log n)$ basic read/write operations to implement each of the snapshot operations and the associated update operation \cite{AttiyaR98}. Such an operation brings a programming comfort as it reduces the ``noise'' introduced by asynchrony and failures \cite{G98} and is particularly used in round-based computations \cite{Gafni98} we consider for the study of the composability of sequential consistency.

\paragraph{Contributions.} We present three major contributions. (1) We identify two contexts that can benefit from the use of sequential 
consistency: round-based algorithms using a different shared object for
each round, and asynchronous shared-memory
systems, where programs can not distinguish a sequentially consistent
memory from a linearizable one.
(2) We propose an implementation of a sequentially consistent memory
where waiting is only required when a write is immediately followed by a
read.
This extends the result presented in~\cite{attiya1994sequential} about
synchronous failure-free systems, to failure-prone asynchronous systems.
(3) The proposed algorithm also implements a sequentially consistent
snapshot operation the cost of which compares very favorably with the
best existing linearizable implementation to our knowledge (the stacking
of the snapshot algorithm of Attiya and Rachman \cite{AttiyaR98} over
the ABD simulation of linearizable registers)

\paragraph{Outline.}
The remainder of this article is organized as follows. In Section~\ref{sec:composition}, we define more formally sequential consistency,
and we present special contexts in which it becomes composable. In Section~\ref{sec:implementation}, we present our implementation 
of shared memory and study its complexity. Finally, Section~\ref{sec:conclusion} concludes the paper.

\section{Sequential Consistency and Composability}\label{sec:composition}

\subsection{Definitions}

In this section we recall the definitions of the most important notions we discuss in this paper: 
two consistency criteria, sequential consistency ($SC$, Def.~\ref{def:SC}, \cite{lamport1979make}) and linearizability ($L$, Def.~\ref{def:Lin}, \cite{herlihy1990linearizability}), 
as well as composability (Def.~\ref{def:comp}). A consistency criterion associates a set of admitted \emph{histories}
to the \emph{sequential specification} of each given object. A history is a representation of an execution. It contains a set of operations, 
that are partially ordered according to the sequential order of each process, called \emph{process order}. 
A sequential specification 
is a language, i.e. a set of sequential (finite and infinite) words. For a consistency criterion $C$ and a sequential specification $T$, 
we say that an algorithm implements a $C(T)$-consistent object if all its executions can be modelled by a history that belongs to $C(T)$, 
that contains all returned operations and only invoked operations. 
Note that this implies that if a process crashes during an operation, then the operation will appear in the history as if it was complete or as if it never took place at all.
\begin{definition}[Linear extension]\label{def:lin_order}
Let $H$ be a history and $T$ be a sequential specification. A \emph{linear extension} $\le$ is a total order on all the operations of $H$, that contains the process order, and such that each event $e$ has a finite past $\{e' : e'\le e\}$ according to the total order. 
\end{definition}
\begin{definition}[Sequential Consistency]\label{def:SC}
    Let $H$ be a history and $T$ be a sequential specification. The history $H$ is \emph{sequentially consistent} regarding $T$, denoted $H\in SC(T)$, if there exists a linear extension $\le$ such that the word composed of all the operations of $H$ ordered by $\le$ belongs to $T$. 
\end{definition}
\begin{definition}[Linearizability]\label{def:Lin}
    Let $H$ be a history and $T$ be a sequential specification. The history $H$ is \emph{linearizable} regarding $T$, 
    denoted $H\in L(T)$, if there exists a linear extension $\le$ such that 
    (1) for two operations $a$ and $b$, if operation $a$ returns before operation $b$ begins, then $a\le b$ and 
    (2) the word formed of all the operations of $H$ ordered by $\le$ belongs to $T$. 
\end{definition}
Let $T_1$ and $T_2$ be two sequential specifications. We define the \emph{composition} of $T_1$ and $T_2$, denoted by $T_1\times T_2$, 
as the set of all the interleaved sequences of a word from $T_1$ and a word from $T_2$. An interleaved sequence of two words $l_1$ and $l_2$ is
a word composed of the disjoint union of all the letters of $l_1$ and $l_2$, that appear in the same order as they appear in $l_1$ and $l_2$. 
For example, the words $ab$ and $cd$ have six interleaved sequences: $abcd$, $acbd$, $acdb$, $cabd$, $cadb$ and $cdab$. 

A consistency criterion $C$ is composable (Def.~\ref{def:comp}) if the composition of a $C(T_1)$-consistent object and a $C(T_2)$-consistent object
is a $C(T_1\times T_2)$-consistent object. Linearizability is composable, and sequential consistency is not. 

\begin{definition}[Composability]\label{def:comp}
For a history $H$ and a sequential specification $T$, let $H_{T}$ be the sub-history of $H$ containing only the operations belonging to $T$. 

A consistency criterion $C$ is \emph{composable} if, for all sequential specifications $T_1$ and $T_2$ and all histories $H$ containing only events 
on $T_1$ and $T_2$, $(H_{T_1} \in C(T_1) \text{ and } H_{T_2} \in C(T_2))$ imply $H \in C(T_1\times T_2)$.
\end{definition}

\subsection{From Linearizability to Sequential Consistency}

\begin{figure}[t]
  \begin{subfigure}{0.17\textwidth}
    \centering
    \scalebox{0.9}{
    \begin{tikzpicture}
      
      \draw[fill=colorB!10] (0.5,2.8) rectangle (2.5,3.6) ;
      \draw[fill=colorB!10] (0.5,1) rectangle (2.5,2.6) ;
      \draw[fill=colorA!10] (0.7,1.2) rectangle (1.4,1.8) ;
      \draw[fill=colorA!10] (1.6,1.2) rectangle (2.3,1.8) ;
      \draw[fill=colorA!10] (0.5,0) rectangle (2.5,0.8) ;

      \draw (1.5,3.2) node{Application};
      \draw (1.5,2.2) node{$Y \times Z$};
      \draw (1.05,1.5) node{$Y$};
      \draw (1.95,1.5) node{$Z$};
      \draw (1.5,0.4) node{$X$ (memory)};

    \end{tikzpicture}
    }
  \caption{Layer based architecture.}
  \label{fig:lin:archi}
  \end{subfigure}
  \hspace{\fill}
  \begin{minipage}{0.8\textwidth} 
  \begin{subfigure}{\textwidth}
    \centering
    \scalebox{0.75}{
    \begin{tikzpicture}
      
      \draw[->] (-0.5,1) node[left]{$p_1$} -- (11.5,1);
      \draw[->] (-0.5,0) node[left]{$p_0$} -- (11.5,0);

      \draw[fill=colorB!10] (0,1) +(2.1,-0.3) rectangle +(10,0.3) +(7,0) node{\footnotesize $Y_{SC}.op_1^0$};

      \draw[fill=colorB!10] (0,0) +(0.2,-0.3) rectangle +(3.5,0.3) +(2.7,0) node{\footnotesize $Y_{SC}.op_0^0$};
      \draw[fill=colorB!10] (0,0) +(4.8,-0.3) rectangle +(11.2,0.3) +(5.6,0) node{\footnotesize $Y_{SC}.op_0^1$};

      \draw[->,colorA,rounded corners, thick]
      (2.5,1.5) -- (3,1.5) -- (3,0.5) --
      (1.1,0.5) -- (1.1,-0.5) --
      (7.9,-0.5) -- (7.9,0.5) --
      (4.9,0.5)  -- (4.9,1.5) --
      (9.1,1.5) -- (9.1,0.5)  --
      (10.3,0.5) -- (10.3,-0.5) -- (10.8,-0.5);

      \draw[fill=colorA!10] (3.0,1) +(-0.7,-0.2) rectangle +(0.7,0.2) +(0,0) node{\footnotesize $X_{SC}.op_1^0$};
      \draw[fill=colorA!10] (4.9,1) +(-0.7,-0.2) rectangle +(0.7,0.2) +(0,0) node{\footnotesize $X_{SC}.op_1^1$};
      \draw[fill=colorA!10] (9.1,1) +(-0.7,-0.2) rectangle +(0.7,0.2) +(0,0) node{\footnotesize $X_{SC}.op_1^2$};

      \draw[fill=colorA!10] (1.1,0) +(-0.7,-0.2) rectangle +(0.7,0.2) +(0,0) node{\footnotesize $X_{SC}.op_0^0$};
      \draw[fill=colorA!10] (7.9,0) +(-0.7,-0.2) rectangle +(0.7,0.2) +(0,0) node{\footnotesize $X_{SC}.op_0^1$};
      \draw[fill=colorA!10] (10.3,0) +(-0.7,-0.2) rectangle +(0.7,0.2) +(0,0) node{\footnotesize $X_{SC}.op_0^2$};
      
    \end{tikzpicture}
    }
  \caption{The implementation of upper layer objects call operations on objects from lower layers.}
  \label{fig:lin:histSC}
  \end{subfigure}

  \begin{subfigure}{\textwidth}
    \centering
    \scalebox{0.75}{
    \begin{tikzpicture}
      
      \draw[->] (-0.5,1) node[left]{$p_1$} -- (11.5,1);
      \draw[->] (-0.5,0) node[left]{$p_0$} -- (11.5,0);

      \draw[fill=colorB!10] (0,1) +(0.3,-0.3) rectangle +(9.3,0.3) +(3.9,0) node{\footnotesize $Y_{SC}.op_1^0$};

      \draw[fill=colorB!10] (0,0) +(0.7,-0.3) rectangle +(3.85,0.3) +(1.5,0) node{\footnotesize $Y_{SC}.op_0^0$};
      \draw[fill=colorB!10] (0,0) +(3.95,-0.3) rectangle +(11.1,0.3) +(7.5,0) node{\footnotesize $Y_{SC}.op_0^1$};

      \draw[->,colorA,rounded corners, thick]
      (0.7,1.5) -- (1.2,1.5) -- (1.2,0.5) --
      (3.0,0.5) -- (3.0,-0.5) --
      (4.8,-0.5) -- (4.8,0.5) --
      (6.6,0.5)  -- (6.6,1.5) --
      (8.4,1.5) -- (8.4,0.5)  --
      (10.2,0.5) -- (10.2,-0.5) -- (10.7,-0.5);

      \draw[fill=colorA!10] (1.2,1) +(-0.7,-0.2) rectangle +(0.7,0.2) +(0,0) node{\footnotesize $X_{SC}.op_1^0$};
      \draw[fill=colorA!10] (6.6,1) +(-0.7,-0.2) rectangle +(0.7,0.2) +(0,0) node{\footnotesize $X_{SC}.op_1^1$};
      \draw[fill=colorA!10] (8.4,1) +(-0.7,-0.2) rectangle +(0.7,0.2) +(0,0) node{\footnotesize $X_{SC}.op_1^2$};

      \draw[fill=colorA!10] (3.0,0) +(-0.7,-0.2) rectangle +(0.7,0.2) +(0,0) node{\footnotesize $X_{SC}.op_0^0$};
      \draw[fill=colorA!10] (4.8,0) +(-0.7,-0.2) rectangle +(0.7,0.2) +(0,0) node{\footnotesize $X_{SC}.op_0^1$};
      \draw[fill=colorA!10] (10.2,0) +(-0.7,-0.2) rectangle +(0.7,0.2) +(0,0) node{\footnotesize $X_{SC}.op_0^2$};
      
    \end{tikzpicture}
    }
  \caption{An asynchronous process cannot differentiate this history from the one on Figure~\ref{fig:lin:histSC}.}
  \label{fig:lin:histL}
  \end{subfigure}
  \end{minipage}
  \caption{In layer based program architecture running on asynchronous systems, local clocks of different processes can be distorted such that it 
  is impossible to differentiate a sequentially consistent execution from a linearizable execution.}
  \label{fig:lin}
\end{figure}
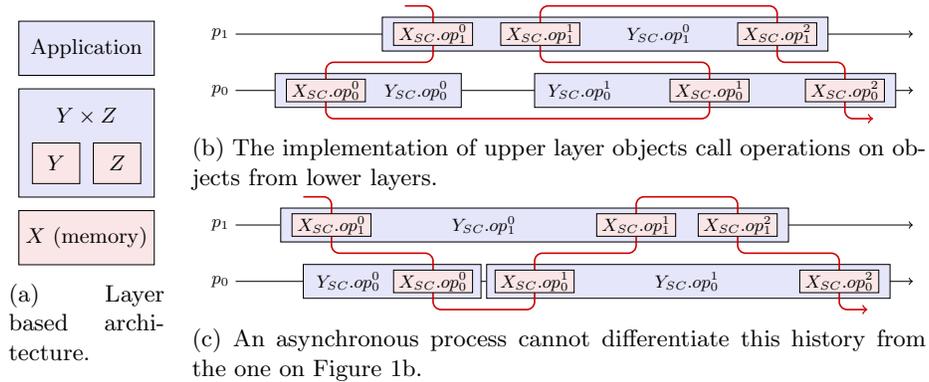

Software developers usually abstract the complexity of their system gradually, which results in a layered software architecture: 
at the top level, an application is built on top of several objects specific to the application, 
themselves built on top of lower levels. Such an architecture is represented in Fig.~\ref{fig:lin:archi}.
The lowest layer usually consists of one or several objects provided by the system itself, typically a shared memory. 
The system can ensure sequential consistency globally on all the provided objects, therefore composability is not required for this level.
Proposition~\ref{prop:lin_SC} expresses the fact that, in asynchronous systems, replacing a linearizable object by a sequentially consistent one 
does not affect the correctness of the programs running on it circumventing the non composability of sequential consistency.
This result may have an impact on parallel architectures, such as modern multi-core processors and, to a higher extent, 
high performance supercomputers, for which the communication with a linearizable central shared memory is very costly, 
and weak memory models such as cache consistency~\cite{goodman1991cache} make the writing of programs tough. 
The idea of the proof is that in any sequentially consistent execution (Fig.~\ref{fig:lin:histSC}), it is possible to associate a local clock 
to each process such that, if these clocks followed real time, the execution would be linearizable (Fig.~\ref{fig:lin:histL}). 
In an asynchronous system, it is impossible for the processes to distinguish between these clocks and real time, so the operations of 
the objects of the upper layers are not affected by the change of clock. 
The complete proof of this proposition can be found in \cite{tech_rep}. 

\begin{proposition}\label{prop:lin_SC}
Let $A$ be an algorithm that implements an $SC(Y)$-consistent object when it is executed on an asynchronous system providing an $L(X)$-consistent object. Then $A$ also implements an $SC(Y)$-consistent object when it is executed in an asynchronous system providing an $SC(X)$-consistent object.
\end{proposition}

An interesting point about Proposition~\ref{prop:lin_SC} is that it allows sequentially consistent --- but not linearizable ---
objects to be composable. Let $A_Y$ and $A_Z$ be two algorithms that implement $L(Y)$-consistent and $L(Z)$-consistent objects
when they are executed on an asynchronous system providing an $L(X)$-consistent object, like on Fig.~\ref{fig:lin:archi}.
As linearizability is stronger than sequential consistency, according to Proposition~\ref{prop:lin_SC}, executing $A_Y$ and $A_Z$
on an asynchronous system providing an $SC(X)$-consistent object would implement sequentially consistent --- yet not linearizable --- 
objects. However, in a system providing the linearizable object $X$, by composability of linearizability, 
the composition of $A_Y$ and $A_Z$ implements an $L(Y\times Z)$-consistent object. Therefore, by Proposition~\ref{prop:lin_SC} again, 
in a system providing the sequentially consistent object $X$, the composition also implements an $SC(Y\times Z)$-consistent object.
In this example, the sequentially consistent versions of $Y$ and $Z$ derive their composability from an anchor to a \emph{common time},
given by the sequentially consistent memory, that can differ from \emph{real time}, required by linearizability.

\subsection{Round-Based Computations}\label{sec:round}

Even at a single layer, a program can use several objects that are not composable, but that are used in a fashion so that the non-composability is invisible to the program. Let us illustrate this with round-based algorithms. The synchronous distributed computing model has been extensively studied and well-understood leading the researchers to try to offer the same comfort when dealing with asynchronous systems, hence the introduction of synchronizers \cite{Awerbuch85}. A synchronizer slices a computation into phases during which each process executes three steps: send/write, receive/read and then local computation. This model has been extended to failure prone systems in the round-by-round computing model \cite{Gafni98} and to the Heard-Of model \cite{CS09} among others. Such a model is particularly interesting when the termination of a given program is only eventual. Indeed, some problems are undecidable in failure prone purely asynchronous systems. In order to circumvent this impossibility, eventually or partially synchronous systems have been introduced \cite{DLS88}. In such systems the termination may hold only after some finite but unbounded time, and the algorithms are implemented by the means of a series of asynchronous rounds each using its own shared objects.

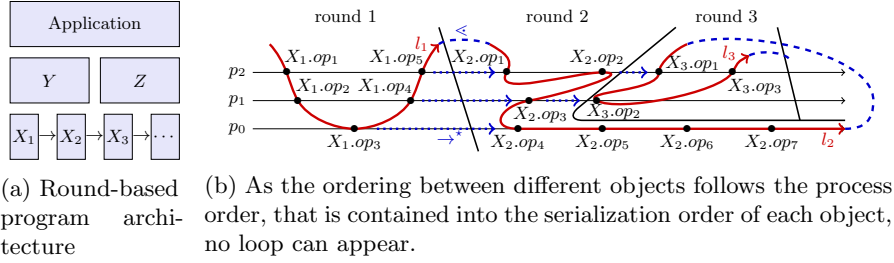
\begin{figure}[t]
  \hspace{\fill}
  \begin{subfigure}{0.2\textwidth}
    \centering
    \scalebox{0.75}{
    \begin{tikzpicture}
      
      \draw[fill=colorB!10] (0,2) rectangle (3,2.8) ;
      \draw[fill=colorB!10] (0,1) rectangle (1.4,1.8) ;
      \draw[fill=colorB!10] (1.6,1) rectangle (3,1.8) ;
      \draw[fill=colorB!10] (0,0) rectangle (0.5,0.8) ;
      \draw[fill=colorB!10] (0.8333,0) rectangle (1.3333,0.8) ;
      \draw[fill=colorB!10] (1.666,0) rectangle (2.166,0.8) ;
      \draw[fill=colorB!10] (2.5,0) rectangle (3,0.8) ;

      \draw (1.5,2.4) node{Application};
      \draw (0.7,1.4) node{$Y$};
      \draw (2.3,1.4) node{$Z$};
      \draw (0.25,0.4) node{$X_1$};
      \draw (0.6666,0.4) node{$\rightarrow$};
      \draw (1.08333,0.4) node{$X_2$};
      \draw (1.5,0.4) node{$\rightarrow$};
      \draw (1.917,0.4) node{$X_3$};
      \draw (2.33333,0.4) node{$\rightarrow$};
      \draw (2.75,0.4) node{$\cdots$};

    \end{tikzpicture}
    }
  \caption{Round-based \\ program architecture}
  \label{fig:rounds:archi}
  \end{subfigure}
  \hspace{\fill}
  \begin{subfigure}{0.75\textwidth}
    \centering
    \scalebox{0.75}{
    \begin{tikzpicture}

      \draw (0.15,2) node{round 1};
      \draw (3.9 ,2) node{round 2};
      \draw (6.9 ,2) node{round 3};

      \draw[thick,rounded corners=10] (1.8,1.8) -- (2.5,-0.4);
      \draw[thick,rounded corners=10] (6,1.8) -- (4,0.15) -- (9,0.15) ;
      \draw[thick,rounded corners=10] (7.8,1.8) -- (8.2,0.15);

      \draw[->] (-1.5,1.0) node[left]{$p_2$} -- (9,1.0) ;
      \draw[->] (-1.5,0.5) node[left]{$p_1$} -- (9,0.5) ;
      \draw[->] (-1.5,0.0) node[left]{$p_0$} -- (9,0.0) ;

      \draw[->,colorA,very thick]
      (-1.2,1.5) to[out=-45,in=110,distance=5]
      (-0.9,1.0) to[out=-70,in=120,distance=5]
      (-0.7,0.5) to[out=-60,in=180,distance=10]
      (0.3,0.0)  to[out=0 ,in=-120,distance=10]
      (1.3,0.5)  to[out=60,in=-110,distance=5]
      (1.5,1.0)  to[out=70,in=-135,distance=5]
      (1.8,1.5);

      \draw[colorB,very thick,dashed]
      (1.8,1.5) to[out=45,in=150,distance=5]
      (2.7,1.5);

      \draw[->,colorA,very thick]
      (2.7,1.5) to[out=-30,in=60,distance=8]
      (3.0,1.0) to[out=-120,in=180,distance=15]
      (4.7,1.0)  to[out=0 ,in=10,distance=17]
      (3.4,0.5)  to[out=-170,in=180,distance=15]
      (3.2,0.0) to[out=0,in=180,distance=20]
      (4.2,0.0) -- (8.2,0.0) -- (9,0.0);

      \draw[colorB,very thick,dashed]
      (9,0) to[out=0,in=-90,distance=10]
      (9.5,.5) to[out=90,in=20,distance=25]
      (6.2,1.5);

      \draw[->,colorA,very thick]
      (6.2,1.5) to[out=-160,in=90,distance=5]
      (5.7,1.0) to[out=-90,in=150,distance=10]
      (4.6,0.5) to[out=-30,in=-120,distance=12]
      (7.0,1.0)  to[out=60,in=-150,distance=7]
      (7.3,1.3);
      
      \draw[colorB,very thick,dashed]
      (7.3,1.3) to[out=30,in=135,distance=5]
      (8,1.25);

      \draw[->, colorB,very thick,dotted] (1.7,1.0) -- (2.8,1.0);
      \draw[->, colorB,very thick,dotted] (1.5,0.5) -- (3.1,0.5);
      \draw[->, colorB,very thick,dotted] (0.7,0.0) -- (2.8,0.0);

      \draw[->, colorB,very thick,dotted] (5,1.0) -- (5.5,1.0);
      \draw[->, colorB,very thick,dotted] (3.7,0.5) -- (4.3,0.5);

      \draw[colorA] (1.5,1.5)  node{$l_1$};
      \draw[colorA] (8.7,-0.2)  node{$l_2$};
      \draw[colorA] (6.95,1.35)  node{$l_3$};
      \draw[colorB] (2.2,1.75)  node{$\lessdot$};
      \draw[colorB] (2,-0.15)  node{$\rightarrow^\star$};

      \draw (-0.9,1.0)node{$\bullet$} +(0.45,0) node[above]{\footnotesize $X_1.op_1$};
      \draw (-0.7,0.5)node{$\bullet$} +(0.45,0) node[above]{\footnotesize $X_1.op_2$};
      \draw (0.3,0.0) node{$\bullet$} node[below]{\footnotesize $X_1.op_3$};
      \draw (1.3,0.5) node{$\bullet$} +(-0.45,0) node[above]{\footnotesize $X_1.op_4$};
      \draw (1.5,1.0) node{$\bullet$} +(-0.45,0) node[above]{\footnotesize $X_1.op_5$};

      \draw (3.0,1.0) node{$\bullet$} +(-0.5,0) node[above]{\footnotesize $X_2.op_1$};
      \draw (4.7,1.0) node{$\bullet$} +(-0.1,0) node[above]{\footnotesize $X_2.op_2$};
      \draw (3.4,0.5) node{$\bullet$} +(0.2,0) node[below]{\footnotesize $X_2.op_3$};
      \draw (3.2,0.0) node{$\bullet$} node[below]{\footnotesize $X_2.op_4$};
      \draw (4.7,0.0) node{$\bullet$} node[below]{\footnotesize $X_2.op_5$};
      \draw (6.2,0.0) node{$\bullet$} node[below]{\footnotesize $X_2.op_6$};
      \draw (7.7,0.0) node{$\bullet$} node[below]{\footnotesize $X_2.op_7$};

      \draw (5.7,1.0) node{$\bullet$} +(0.6,-0.1) node[above]{\footnotesize $X_3.op_1$};
      \draw (4.6,0.5) node{$\bullet$} +(0.3,0.05) node[below]{\footnotesize $X_3.op_2$};
      \draw (7.0,1.0) node{$\bullet$} +(0.4,0) node[below]{\footnotesize $X_3.op_3$};
      
    \end{tikzpicture}
    }
  \caption{As the ordering between different objects follows the process order, that is contained into the 
  serialization order of each object, no loop can appear.}
  \label{fig:rounds:hist}
  \end{subfigure}
  \hspace{\fill}
  \caption{The composition of sequentially consistent objects used in different rounds is sequentially consistent.}
  \label{fig:rounds}
\end{figure}

In the round-based computing model the execution is sliced into a sequence of asynchronous rounds. During each round, a new data structure (usually a single-writer/multi-reader register per process) is created and it is the only shared object used to communicate during the round. 
At the end of the round, each process destroys its local accessor to the object, so that it can no more access it. Note that the rounds are asynchronous:
the processes do not necessarily start and finish their rounds at the same time. Moreover, a process may not terminate a round and keep accessing the same shared object forever or may crash during this round and stop executing. A round-based execution is illustrated in Fig.~\ref{fig:rounds:hist}.

In Proposition~\ref{prop:round}, we prove that sequentially consistent objects of different rounds behave well together:
as the ordering added between the operations of two different objects always follows the round numbering, that is 
consistent with the program order already contained in the linear extension of each object, the composition of 
all these objects cannot create loops (Figure~\ref{fig:rounds:hist}). The complete proof of this 
proposition can be found in \cite{tech_rep}. 
Putting together this result and Proposition~\ref{prop:lin_SC}, all the algorithms 
that use a round-based computation model can benefit of any improvement on the implementation of an array of 
single-writer/multi-reader register that sacrifices linearizability for sequential consistency. 
Note that this remains true whatever is the data structure used during each round. 
The only constraint is that a sequentially consistent shared data structure can be accessed during a unique round. 
If each object is sequentially consistent then the whole execution is consistent.

\begin{proposition}\label{prop:round}
  Let $(T_r)_{r\in\mathbb{N}}$ be a family of sequential specifications and $(X_r)_{r\in\mathbb{N}}$ be a 
  family of shared objects such that, for all $r$, $X_r$ is $SC(T_r)$-consistent. Let $H$ be a history
  that does not contain two operations $X_r.a$ and $X_{r'}.b$ with $r>r'$ such that $X_r.a$ precedes $X_{r'}.b$ in the process order. 
  Then $H$ is sequentially consistent with respect to the composition of all the $T_r$.
\end{proposition}

\section{Implementation of a Sequentially Consistent Memory}\label{sec:implementation}
\subsection{Computation Model}\label{sec:model}
The computation system consists of a set $\Pi$ of $n$ sequential processes, denoted $p_0, p_1, \ldots, p_{n-1}$. 
The processes are asynchronous, in the sense that they all proceed at their own speed, 
not upper bounded and unknown to all other processes. 

Among these $n$ processes, up to $t$ may crash (halt prematurely) but otherwise execute correctly the algorithm until the moment of their crash. 
We call a process \emph{faulty} if it crashes, otherwise it is called \emph{correct} or \emph{non-faulty}. 
In the rest of the paper we will consider the above model restricted to the case $t<\frac{n}{2}$.

The processes communicate with each other by sending and receiving messages through a complete network of bidirectional channels. 
A process can directly communicate with any other process, including itself ($p_i$ receives its own messages instantaneously), 
and can identify the sender of the message received. Each process is equipped with two operations: \textbf{send} and \textbf{receive}. 

The communication channels are reliable (no losses, no creation, no duplication, no alteration of messages) and asynchronous 
(finite time needed for a message to be transmitted but there is no upper bound). We also assume 
the channels are FIFO: if $p_i$ sends two messages to $p_j$, $p_j$ will receive them in the order they were sent.
As stated in \cite{birman1987reliable}, FIFO channels can always be implemented on top of non-FIFO channels. 
Therefore, this assumption does not bring additional computational power to the model, but it allows us to simplify the writing of the algorithm. Process $p_i$ can also use the macro-operation \textbf{FIFO broadcast}, that can be seen as a multi-send that sends a message to all processes, including itself. Hence, if a faulty process crashes during the broadcast operation some processes may receive the message while others may not, otherwise all correct processes will eventually receive the message.

\subsection{Single-Writer/Multi-Reader Registers and Snapshot Memory}\label{sec:memory}
The shared memory considered in this paper, called a \emph{snapshot memory}, consists of an array of shared registers denoted $\REG[1..n]$. Each entry $\REG[i]$ represents a single-writer/multi-reader (SWMR) register. When process $p_i$ invokes $\REG.\AlgoWrite(v)$, the value $\Algov$ is written into the SWMR register $\REG[i]$ associated with process $p_i$. Differently, any process $p_i$ can read the whole array $\REG$ by invoking a single operation namely $\REG.\AlgoSnap()$. According to the sequential specification of the snapshot memory, $\REG.\AlgoSnap()$ returns an array containing the most recent value written by each process or the initial default value if no value is written on some register. 
Concurrency is possible between snapshot and writing operations, as soon as the considered consistency criterion, namely linearizability or sequential consistency, is respected. Informally in a sequentially consistent snapshot memory, each snapshot operation must return the last value written by the process that initiated it,
and for any pair of snapshot operations, one must return values at least as recent as the other for all registers.

Compared to read and write operations, the snapshot operation is a higher level abstraction introduced in \cite{Afek93} that eases program design without bringing additional power with respect to shared registers. Of course this induces an additional cost: the best known simulation, above SWMR registers proposed in \cite{AttiyaR98}, needs $O(n\log n)$ basic read/write operations to implement each of the snapshot and the associated update operations.

Since the seminal paper \cite{attiya1995sharing} that proposed the so-called ABD simulation that emulates a linearizable shared memory over a message-passing distributed system, most of the effort has been put on the shared memory model given that a simple stacking allows to translate any shared memory-based result to the message-passing system model. 
Several implementations of linearizable snapshot have been proposed in the literature some works consider variants of snapshot (e.g. immediate snapshot \cite{BorowskyG92}, weak-snapshot \cite{dwork1992time}, one scanner \cite{KirousisST94}) others consider that special constructions such as test-and-set (T\&S) \cite{AHR95} or load-link/store-conditional (LL/SC) \cite{RST01} are available, the goal being to enhance time and space efficiency.
In this paper, we propose the first message-passing sequentially consistent (not linearizable) snapshot memory implementation directly over a message-passing system (and consequently the first sequentially consistent array of SWMR registers), as traditional read and write operations can be immediately deduced from snapshot and update with no additional cost.

\subsection{The Proposed Algorithm}\label{sec:algo}
Algorithm~\ref{algo:SCS} proposes an implementation of the sequentially
consistent snapshot memory data structure presented in
Section~\ref{sec:memory}. The complete proof of correctness
of this algorithm can be found in the technical report~\cite{tech_rep}.
Process $p_i$ can write a value $\Algov$ in its own register $\REG[i]$
by calling the operation $REG.\AlgoWrite(v)$
(lines~\ref{al:SCS:w1}-\ref{al:SCS:w4}). It can also call the operation
$REG.\AlgoSnap()$ (lines~\ref{al:SCS:r1}-\ref{al:SCS:r2}).
Roughly speaking, the principle of this algorithm is to maintain, on
each process, a local view of the object that reflects a set of
\emph{validated} update operations.
To do so, when a value is written, all processes label it with their own
timestamp. The order in which processes timestamp two different update
operations define a \emph{dependency relation} between these operations.
For two operations $a$ and $b$, if $b$ depends on $a$, then $p_i$ cannot
validate $b$ before $a$.

More precisely, each process $p_i$ maintains five local variables:
\begin{itemize}
\item $\AlgoX_i \in \mathbb{N}^n$ is the array of most recent validated values written on each register.
\item $\AlgoVC_i \in \mathbb{N}^n$ represents the timestamps associated with the values stored in $\AlgoX_i$, labelled by the process that initiated them.
\item $\AlgoSC_i\in \mathbb{N}$ is an integer clock used by $p_i$ to timestamp all the update operations. $\AlgoSC_i$ is incremented each time a message is sent,
    which ensures all timestamps from the same process are different. 
\item $\AlgoG_i \subset \mathbb{N}^{3+n}$ encodes the dependencies between update
    operations that have not been validated yet, as known by $p_i$. An element $\Algog\in \AlgoG_i$, of the form $(\Algog.\AlgoGV, \Algog.\AlgoGK, \Algog.\AlgoGT, \Algog.\AlgoGCL)$, represents
    the update operation of value $\Algog.\AlgoGV$ by process $p_{\Algog.\AlgoGK}$ labelled by process $p_{\Algog.\AlgoGK}$ with timestamp $\Algog.\AlgoGT$. For all $0\leq j<n$,
    $\Algog.\AlgoGCL[j]$ contains the timestamp given by $p_j$ if it is known by $p_i$, and $\infty$ otherwise. 
    
    All updates of a history can be uniquely represented by a pair of integers $(k, t)$, where $p_k$ is the process that invoked it, and $t$ is the timestamp associated 
    to this update by $p_k$. Considering a history and a process $p_i$, we define the dependency relation $\rightarrow_i$ on pairs of integers $(k, t)$, by 
    $(k, t)\rightarrow_i (k', t')$ if for all $\Algog, \Algog'$ ever inserted in $G_i$ with
    $(\Algog.\AlgoGK,\Algog.\AlgoGT)=(k,t)$, $(\Algog'.\AlgoGK,\Algog'.\AlgoGT)=(k',t')$, 
    we have $|\{j : \Algog'.\AlgoGCL[j] < \Algog.\AlgoGCL[j] \}| \le \frac{n}{2}$ (i.e. the dependency does not exist if $p_i$ knows that a majority of processes have seen the first update before the second).
    Let $\rightarrow_i^\star$ denote the transitive closure of $\rightarrow_i$.
\item $\AlgoV_i\in \mathbb{N}\cup\{\bot\}$ is a buffer register used to store a value written while the previous one is not yet validated. This is necessary for validation (see below).
\end{itemize}

The key of the algorithm is to ensure the inclusion between sets of validated updates on any two processes at any time. Remark that it is not always necessary 
to order all pairs of update operations to implement a sequentially consistent snapshot memory: for example, two update operations on different registers commute. 
Therefore, instead of validating both operations on all processes in the exact same order (which requires Consensus), we can validate them at the same 
time to prevent a snapshot to occur between them. Thus, it is sufficient to ensure that, for all pairs of update operations, there is a dependency 
agreed by all processes (possibly in both directions). This is expressed by Lemma~\ref{lemma:safety}.

\begin{algorithm}[t!]
  \scriptsize
  \tcc{Local variable initialization}
  $\KwX_i  \leftarrow [0,\dots, 0]$\label{al:SCS:varX}\tcp*{$\KwX_i[j]\in \mathbb{N}$: last validated value written by $p_j$}
  $\KwVC_i \leftarrow [0,\dots, 0]$\label{al:SCS:varVC}\tcp*{$\KwVC_i[j] \in \mathbb{N}$: stamp given by $p_j$ to $\KwX_i[j]$}
  $\KwSC_i \leftarrow 0$\label{al:SCS:varSC}\tcp*{used to stamp all the updates}
  $\KwG_i  \leftarrow \emptyset$\label{al:SCS:varG}\tcp*{contains a $g = (g.\KwGV, g.\KwGK, g.\KwGT, g.\KwGCL)$ per non-val. update}
  $\KwV_i  \leftarrow \bot$\label{al:SCS:varV}\tcp*{$\KwV_i \in \mathbb{N}\cup \{\bot\}$: stores postponed updates}

  \nonl\hrulefill\\
  \nonl\SubAlgo{\Op $\KwWrite(\Kwv)$  \tcc*[h]{$\Kwv\in \mathbb{N}$: written value; no return value}}{
    \uIf(\label{al:SCS:w1}\tcp*[f]{no non-validated update by $p_i$}){$\forall g\in \KwG_i: g.\KwGK\neq i$} {
      $\KwSC_i\text{++}$\; \Broadcast $\KwM(\Kwv,i,\KwSC_i,\KwSC_i)$\;\label{al:SCS:w3}
    }
    \lElse(\label{al:SCS:w4}\tcp*[f]{postpone the update}){$\KwV_i \leftarrow \Kwv$}
  }

  \nonl\hrulefill\\
  \nonl\SubAlgo{\Op $\KwSnap()$  \tcc*[h]{return type: $\mathbb{N}^n$}}{
    \Wait{$\KwV_i=\bot \land \forall g\in \KwG_i: g.\KwGK\neq i$} \label{al:SCS:r1}\tcp*[r]{make sure $p_i$'s updates are validated}
    \Return $\KwX_i$\label{al:SCS:r2}\;
  }

  \nonl\hrulefill\\
  \nonl\SubAlgo{\RecA $\KwM(\Kwv,\Kwk,\Kwt,\Kwcl)$ \RecB $p_j$} {
    \tcp{$\Kwv\in \mathbb{N}$: written value, $\Kwk\in \mathbb{N}$: writer id, $\Kwt\in \mathbb{N}$: stamp by $p_{\Kwk}$, $\Kwcl\in \mathbb{N}$: stamp by $p_j$}
    \If(\label{al:SCS:mA1}\tcp*[f]{update not validated yet}){$\Kwt>\KwVC_i[\Kwk]$} {
      \eIf(\label{al:SCS:mA2}\tcp*[f]{update already known}){$\exists g\in \KwG_i: g.\KwGK = \Kwk \land g.\KwGT = \Kwt$}{
        $g.\KwGCL[j] \leftarrow \Kwcl$\;\label{al:SCS:mA3}
      }(\label{al:SCS:mA4}\tcp*[f]{first message for this update}){
        \If{$\Kwk\neq i$}{
          $\KwSC_i\text{++}$ \tcp*[r]{forward with own stamp}
          \Broadcast $\KwM(\Kwv, \Kwk, \Kwt, \KwSC_i)$\label{al:SCS:mA6}\;
        }
        \textbf{var} $g \leftarrow \left(g.\KwGV = \Kwv, g.\KwGK=\Kwk, g.\KwGT=\Kwt, g.\KwGCL=[\infty,\dots,\infty]\right)$\label{al:SCS:mA7};
        $g.\KwGCL[j] \leftarrow \Kwcl$\label{al:SCS:mA8}\;
        $\KwG_i \leftarrow \KwG_i\cup \{g\}$\label{al:SCS:mA9}\tcp*[r]{create an entry in $\KwG_i$ for the update}
      }
    }
    \textbf{var} $G' = \{g \in \KwG_i: |\{l : g'.\KwGCL[l] < \infty\}| > \frac{n}{2} \} $\label{al:SCS:mB1}\tcp*[r]{$G'$ contains validable updates}
    \lWhile(\label{al:SCS:mB2}){$\exists g\in \KwG_i\setminus G', g'\in G' : |\{l : g'.\KwGCL[l] < g.\KwGCL[l] \}| \neq \frac{n}{2}$}{
      $G'\leftarrow G'\setminus \{g'\}$\label{al:SCS:mB3}}
    $\KwG_i \leftarrow \KwG_i \setminus G'$\label{al:SCS:mB4}\tcp*[r]{validate updates of $G'$}
    \For(\label{al:SCS:mB5}){$g\in G'$}{
      \lIf{$\KwVC_i[g.\KwGK]<g.\KwGT$}{$\KwVC_i[g.\KwGK]=g.\KwGT;$ $\KwX_i[g.\KwGK]=g.\KwV$}\label{al:SCS:mB6}
    }
    \If(\label{al:SCS:mC1}\tcp*[f]{start validation process for}){$\KwV_i\neq\bot \land \forall g\in \KwG_i: g.\KwGK\neq i$}{
      $\KwSC_i\text{++}$ \tcp*[r]{postponed update if any}
      \Broadcast $\KwM(\KwV_i,i,\KwSC_i,\KwSC_i)$\label{al:SCS:mC3}\;
      $\KwV_i \leftarrow \bot$\label{al:SCS:mC4}\;
    }
  }
  \caption{Implementation of a sequentially consistent memory (for $p_i$)}
  \label{algo:SCS}
\end{algorithm}

\begin{lemma} \label{lemma:safety}
Let $p_i$, $p_j$ be two processes and $t_i$, $t_j$ be two time instants, and let us denote by $\AlgoVC_i^{t_i}$ (resp. $\AlgoVC_j^{t_j}$) the value of $\AlgoVC_i$ (resp. $\AlgoVC_j$) at time $t_i$ (resp. $t_j$). We have either, for all $k$, $\AlgoVC_i^{t_i}[k] \le \AlgoVC_j^{t_j}[k]$ or for all $k$, $\AlgoVC_j^{t_j}[k] \le \AlgoVC_i^{t_i}[k]$.
\end{lemma}

This is done by the mean of messages of the form $\AlgoM(\Algov,\Algok,\Algot,\Algocl)$ containing four integers: $\Algov$ the value written, $\Algok$ the identifier of the process that initiated the update, $\Algot$ the timestamp given by $p_{\Algok}$ and $\Algocl$ the timestamp given by the process that sent this message. Timestamps 
of successive messages sent by $p_i$ are unique and totally ordered, thanks to variable $\AlgoSC_i$, that is incremented each time a message is sent by $p_i$.
When process $p_i$ wants to submit a value $\Algov$ for validation, it FIFO-broadcasts a message $\AlgoM(\Algov,i,\AlgoSC_i,\AlgoSC_i)$ (lines~\ref{al:SCS:w3} and~\ref{al:SCS:mC3}). 
When $p_i$ receives a message $\AlgoM(\Algov,\Algok,\Algot,\Algocl)$, three cases are possible. If $p_i$ has already validated the corresponding update ($\Algot > \AlgoVC_i[\Algok]$),
the message is simply ignored. Otherwise, if it is the first time $p_i$ receives a message concerning this update ($\AlgoG_i$ does not contain any piece of information concerning it), 
it FIFO-broadcasts a message with its own timestamp and adds a new entry $\Algog\in \AlgoG_i$. Whether it is its first message or not, $p_i$ records the timestamp $\Algocl$, given by $p_j$, in $\Algog.\AlgoGCL[j]$ (lines~\ref{al:SCS:mA3} or~\ref{al:SCS:mA8}). Note that we cannot update $\Algog.\AlgoGCL[\Algok]$ at this point, as the broadcast is not causal: if $p_i$ did so, it could miss dependencies imposed by the order in which $p_{\Algok}$ saw concurrent updates. Then, $p_i$ tries to validate update operations: $p_i$ can validate an operation $a$ if it has received messages from a majority of processes, and there is no operation $b\rightarrow_i^\star a$ that cannot be validated. For that, it creates the set $G'$ that initially contains all the operations that have received enough messages, and removes all operations with unvalidatable dependencies from it (lines~\ref{al:SCS:mB1}-\ref{al:SCS:mB3}), and then updates $\AlgoX_i$ and $\AlgoVC_i$ with the most recent validated values (l
 ines~\ref{al:SCS:mB4}-\ref{al:SCS:mB6}).

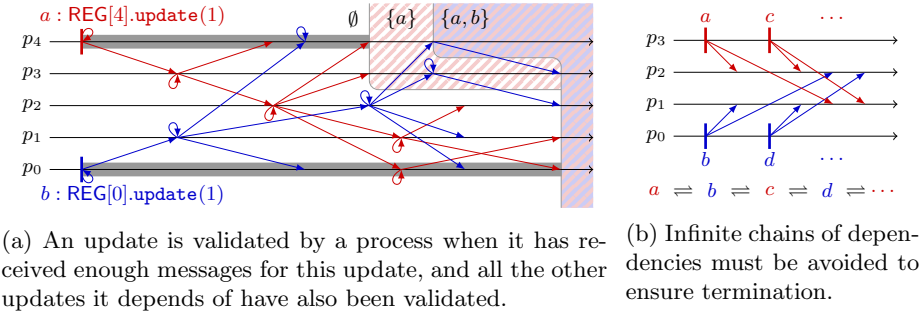
\begin{figure}[t]
  \begin{subfigure}{.66\textwidth}
    \centering
    \scalebox{0.85}{
      \begin{tikzpicture}

        \fill[colorB!20] (5,2.6) {[rounded corners=5] -- (5,1.75) -- (7,1.75)} -- (7,-0.6) -- (7.5,-0.6) -- (7.5,2.6) -- cycle;
        \fill[pattern=stripes, pattern color=colorA!20] (4,2.6) {[rounded corners=5] -- (4,1.25)} -- (7,1.25) -- (7,-0.6) -- (7.5,-0.6) -- (7.5,2.6) -- cycle;
        \draw[black!40,rounded corners] (5,2.6) -- (5,1.75) -- (7,1.75) -- (7,-0.6);
        \draw[black!40,rounded corners] (4,2.6) -- (4,1.25) -- (7,1.25);
        \draw[draw=black!40,fill=black!40] (-0.5,-0.1) rectangle (7,0.1);
        \draw[draw=black!40,fill=black!40] (-0.5,1.9) rectangle (4,2.1);

        \draw[->] (-1,2.0) node[left]{$p_4$} -- (7.5,2.0);
        \draw[->] (-1,1.5) node[left]{$p_3$} -- (7.5,1.5);
        \draw[->] (-1,1.0) node[left]{$p_2$} -- (7.5,1.0);
        \draw[->] (-1,0.5) node[left]{$p_1$} -- (7.5,0.5);
        \draw[->] (-1,0.0) node[left]{$p_0$} -- (7.5,0.0);

        \draw (3.75,2.35) node{$\emptyset$};
        \draw (4.5,2.35) node{$\{a\}$};
        \draw (5.5,2.35) node{$\{a, b\}$};

        \draw[colorA,  very thick] (-0.5,2.0) +(0,-0.2) -- +(0,0.2) +(0.8,0.4) node{$a: \REG[4].\AlgoWrite(1)$};
        \draw[colorB, very thick] (-0.5,0.0) +(0,-0.2) -- +(0,0.2) +(0.8,-0.4) node{$b: \REG[0].\AlgoWrite(1)$};

        \draw[-latex, colorA] (-0.5,2.0) to[out=90,in=30,distance=10] (-0.5,2.0);
        \draw[-latex, colorA] (-0.5,2.0) -- (1,1.5);
        \draw[-latex,colorA] (1,1.5) to[out=-150,in=-90,distance=10] (1,1.5);
        \draw[-latex,colorA] (1,1.5) -- (2.5,1.0);
        \draw[-latex,colorA] (1,1.5) -- (2.5,2.0);
        \draw[-latex,colorA] (2.5,1.0) to[out=-150,in=-90,distance=10] (2.5,1.0);
        \draw[-latex,colorA] (2.5,1.0) -- (4.0,1.5);
        \draw[-latex,colorA] (2.5,1.0) -- (4.0,2.0);
        \draw[-latex,colorA] (2.5,1.0) -- (4.5,0.5);
        \draw[-latex,colorA] (2.5,1.0) -- (4.5,0.0);
        \draw[-latex,colorA] (4.5,0.5) to[out=-150,in=-90,distance=10] (4.5,0.5);
        \draw[-latex,colorA] (4.5,0.5) -- (7.0,0.0);
        \draw[-latex,colorA] (4.5,0.5) -- (5.50,1.0);
        \draw[-latex,colorA] (4.5,0.0) to[out=-150,in=-90,distance=10] (4.5,0.0);
        \draw[-latex,colorA] (4.5,0.0) -- (7.0,0.5);

        \draw[-latex,colorB] (-0.5,0.0) to[out=-90,in=-30,distance=10] (-0.5,0.0);
        \draw[-latex,colorB] (-0.5,0.0) -- (1,0.5);
        \draw[-latex,colorB] (1.0,0.5) to[out=150,in=90,distance=10] (1.0,0.5);
        \draw[-latex,colorB] (1.0,0.5) -- (4.0,1.0);
        \draw[-latex,colorB] (1.0,0.5) -- (3.0,0.0);
        \draw[-latex,colorB] (1.0,0.5) -- (3.0,2.0);
        \draw[-latex,colorB] (4.0,1.0) to[out=150,in=90,distance=10] (4.0,1.0);
        \draw[-latex,colorB] (4.0,1.0) -- (5.5,0.0);
        \draw[-latex,colorB] (4.0,1.0) -- (5,1.5);
        \draw[-latex,colorB] (4.0,1.0) -- (5,2.0);
        \draw[-latex,colorB] (4.0,1.0) -- (5.5,0.5);
        \draw[-latex,colorB] (5.0,1.5) to[out=150,in=90,distance=10] (5.0,1.5);
        \draw[-latex,colorB] (5.0,1.5) -- (7.0,1.0);
        \draw[-latex,colorB] (3,2.0) to[out=150,in=90,distance=10] (3,2.0);
        \draw[-latex,colorB] (5.0,2.0) -- (7.0,1.5);

      \end{tikzpicture}
    }
    \caption{An update is validated by a process when it has received enough messages for this update, and all the other updates it depends of have also been validated.}
    \label{fig:expl_algo:handshake}
  \end{subfigure}
  \hspace{\fill}
  \begin{subfigure}{.32\textwidth}
    \centering
    \scalebox{0.85}{
      \begin{tikzpicture}

        \colorlet{colorA}{black!20!red}
        \colorlet{colorB}{black!20!blue}
        
        \draw[colorA] (0.2,-0.85) node{$a$};
        \draw         (0.65,-0.85) node{$\rightleftharpoons$};
        \draw[colorB] (1.1,-0.85) node{$b$};
        \draw         (1.55,-0.85) node{$\rightleftharpoons$};
        \draw[colorA] (2.0,-0.85) node{$c$};
        \draw         (2.45,-0.85) node{$\rightleftharpoons$};
        \draw[colorB] (2.9,-0.85) node{$d$};
        \draw         (3.35,-0.85) node{$\rightleftharpoons$};
        \draw[colorA] (3.8,-0.85) node{$\dots$};
        
        \draw[->] (0.5,1.5) node[left]{$p_3$} -- (4,1.5);
        \draw[->] (0.5,1.0) node[left]{$p_2$} -- (4,1.0);
        \draw[->] (0.5,0.5) node[left]{$p_1$} -- (4,0.5);
        \draw[->] (0.5,0.0) node[left]{$p_0$} -- (4,0.0);

        \draw[colorA, very thick] (1,1.5) +(0,-0.2) -- +(0,0.2) +(0,0.35) node{$a$};
        \draw[colorA, very thick] (2,1.5) +(0,-0.2) -- +(0,0.2) +(0,0.35) node{$c$};
        \draw[colorA] (3,1.85) node{$\dots$};

        \draw[colorB, very thick] (1,0.0) +(0,-0.2) -- +(0,0.2) +(0,-0.35) node{$b$};
        \draw[colorB, very thick] (2,0.0) +(0,-0.2) -- +(0,0.2) +(0,-0.35) node{$d$};
        \draw[colorB] (3,-0.35) node{$\dots$};

        \draw[-latex,colorA] (1,1.5) -- (1.5,1.0);
        \draw[-latex,colorA] (2,1.5) -- (2.5,1.0);
        \draw[-latex,colorB] (1,0.0) -- (3,1.0);
        \draw[-latex,colorB] (2,0.0) -- (3.5,1.0);

        \draw[-latex,colorB] (1,0.0) -- (1.5,0.5);
        \draw[-latex,colorB] (2,0.0) -- (2.5,0.5);
        \draw[-latex,colorA] (1,1.5) -- (3,0.5);
        \draw[-latex,colorA] (2,1.5) -- (3.5,0.5);
      \end{tikzpicture}
    }
    \caption{Infinite chains of dependencies must be avoided to ensure termination.}
    \label{fig:expl_algo:dependences}
    \end{subfigure}
    \caption{Two executions of Algorithm~\ref{algo:SCS}}
    \label{fig:expl_algo}
\end{figure}

This mechanism is illustrated in Fig. \ref{fig:expl_algo:handshake}.
Processes $p_0$ and $p_4$ initially call operation $\REG.\AlgoWrite(1)$.
Messages that have an impact in the algorithm are depicted by arrows and
messages that do not appear are received later.
The simplest case is process $p_3$ that received three messages
concerning $a$ (from $p_4$, $p_3$ and $p_2$, with $3>\frac{n}{2}$)
before its first message concerning $b$, allowing it to validate $a$.
The case of $p_4$ is similar: even if it knows that process $p_1$ saw
$b$ before $a$, it received messages concerning $a$ from three
\emph{other} processes, which allows it to ignore the message from $p_1$.
The situation of $p_0$ and $p_1$ may look similar to this of $p_4$, but
the message they received concerning $a$
and one of the messages they received concerning $b$ are from the same
process $p_2$, forcing them to respect the dependency $a\rightarrow_0 b$.
The same situation occurs for $p_2$ so even if $a$ was validated before
$b$ by other processes, $p_2$ must respect the dependency
$b\rightarrow_2 a$.

Sequential consistency requires the total order to contain the process
order. Therefore, a snapshot of process $p_i$ must return values at
least as recent as its last updated value, i.e. it is not allowed to
return from a snapshot between an update and the time of its validation
(grey zones in Fig.~\ref{fig:expl_algo:handshake}). This can be done in
two ways: either by waiting at the end of each update until it is
validated, in
which case all snapshot operations are done for free, or by waiting at
the beginning of all snapshots that immediately follow an update.
This extends the remark of~\cite{attiya1994sequential} to crash-prone
asynchronous systems: to implement a sequentially consistent memory it
is necessary and
sufficient to wait during either read or write operations. In
Algorithm~\ref{algo:SCS}, we chose to wait during read/snapshot
operations (line~\ref{al:SCS:r1}).
This is more efficient for two reasons: first, it is not needed to wait
between two consecutive updates, which cannot be avoided in the other
case, and second the time between the end of an update and the beginning
of a snapshot counts in the validation
process, but it can be used for local computations. Note that when two
snapshot operations are invoked successively, the second one also
returns immediately,
which improves the result of~\cite{attiya1994sequential} according to
which waiting is necessary for all the operations of one kind.

In order to obtain termination of the snapshot operations (and progress
in general), we must ensure that all update operations are eventually
validated by all processes. This is expressed by Lemma~\ref{lemma:liveness}.
Figure \ref{fig:expl_algo:dependences} shows such a case.
Process $p_2$ receives a message concerning $a$ and a message concerning
$c$ before a message concerning $b$, while
 $p_1$ receives a message concerning $b$ before messages concerning $a$
and $c$. This may create dependencies
$a\rightarrow_i b \rightarrow_i c \rightarrow_i b \rightarrow_i a$ on a
process $p_i$  thus forcing $p_i$ to validate $a$ and $c$
at the same time, even if they are ordered by the process order.
Fig.~\ref{fig:expl_algo:dependences} shows that it can
result in an infinite chain of dependencies, blocking validation of any
update operation. To break this chain, we force process $p_3$ to wait
until $a$ is validated
locally before it proposes $c$ to validation by storing the value
written by $c$ in a local variable $\AlgoV_i$ until $a$ is validated
(lines~\ref{al:SCS:w1} and~\ref{al:SCS:w4}). When $a$ is validated,
we start the same validation process for $c$
(lines~\ref{al:SCS:mC1}-\ref{al:SCS:mC4}).
Note that, if several updates (say $c$ and $e$) happen before $a$ is
validated, the update of $c$ can be dropped as it will eventually be
overwritten by $e$.
In this case, $c$ will happen just before $e$ in the final linearization
required for sequential consistency.

\begin{lemma} \label{lemma:liveness}
If a message $\AlgoM(\Algov,i,\Algot,\Algot)$ is sent by a correct process $p_i$, then beyond some time $t'$, 
for each correct process $p_j$, $\AlgoVC_j^{t'}[i] \ge \Algot$.
\end{lemma}

We can now prove that all histories 
admitted by Algorithm \ref{algo:SCS} are sequentially consistent with respect to the snapshot memory object. 
The idea is to order snapshot operations according to the order given by Lemma~\ref{lemma:safety} on
the value of $\AlgoVC_i$ when they were made and to insert the update operations at the position where 
$\AlgoVC_i$ changes because they are validated. This order can be completed into a linear 
extension, by Lemma~\ref{lemma:liveness}, and to show that the execution of all the operations in that order
respects the sequential specification of the snapshot memory data structure.
The complete proof can be found in \cite{tech_rep}. 


\subsection{Complexity}\label{sec:complexity}

In this section, we analyze the complexity of Algorithm~\ref{algo:SCS}
in terms of number of messages
and latency for each operation. We compare the complexity of our
algorithm with the standard implementation of
linearizable registers in~\cite{attiya1995sharing} with unbounded
messages. Note that~\cite{attiya1995sharing} also proposes an
implementation with bounded messages but at a much higher cost in terms of
latency, which is the parameter we are really interested in improving in
this paper. As our algorithm also implements the snapshot operation,
we compare it to the implementation of a snapshot object
\cite{AttiyaR98} on top of registers. Fig.~\ref{fig:complexity} sums up
these complexities.

We measure the complexity as the length of the longest chain of causally related messages to expect before an operation can complete, e.g. 
if a process sends a message and then waits for some answers, the complexity will be $2$.

Each update generates at most $n^2$ messages and has latency $0$, as
update operations return immediately.
No message is sent for snapshot operations. In terms of latency, in the
worst case a snapshot
is called directly after two update operations $a$ and $b$: the process
must wait for acknowledgements for its message for $a$, and then for acknowledgements 
for its message for $b$, which gives a complexity of $4$. 
However, if enough time has elapsed between a snapshot and the last update, the snapshot returns immediately.

In comparison, the ABD simulation uses solely a linear number of messages per operation (reads as well as writes), but waiting is necessary for both 
kinds of operations. Even in the case of the read operation, our worst case corresponds to the latency of the ABD simulation. 
Moreover, our solution directly implements the snapshot operation. Implementing a snapshot operation on top of a linearizable shared memory is in fact more costly than just reading each register once. The AR implementation \cite{AttiyaR98}, that is (to our knowledge) 
the implementation of the snapshot that uses the least amount of operations on the registers, uses $\mathcal{O}(n\log n)$ operations on registers
to complete both a snapshot and an update operation. As each operation on memory 
requires $\mathcal{O}(n)$ messages and has a latency of $\mathcal{O}(1)$, our approach leads to a better performance in all cases. 

\begin{figure}[t]
    \centering
    \scalebox{0.69}{
    \begin{tikzpicture}

      \draw (-0.1,0) -- (17.3,0) ;
      \draw (-0.1,0.5) -- (17.3,0.5) ;
      \draw (-0.1,1) -- (17.3,1) ;
      \draw (-0.1,1.5) -- (17.3,1.5) ;
      \draw (2.6,2.4) -- (17.3,2.4) ;

      \draw (-0.1,0) -- (-0.1,1.5) ;

      \draw (1.2,1.25) node{ABD \cite{attiya1995sharing}} ;
      \draw (1.2,0.75) node{ABD + AR \cite{attiya1995sharing,AttiyaR98}} ;
      \draw (1.2,0.25) node{Algorithm \ref{algo:SCS}} ;

      \draw (2.5,0) -- (2.5,1.5) ;
      \draw (2.6,0) -- (2.6,2.4) ;

      \draw (4.15,2.15) node{Read} ;

      \draw (3.5,1.75) node{\# messages} ;
      \draw (3.5,1.25) node{$\mathcal{O}(n)$} ;
      \draw[black!50] (3.5,0.75) node{$\sim$} ;
      \draw (3.5,0.25) node{$0$} ;

      \draw (4.4,0) -- (4.4,1.5) ;

      \draw (5.05,1.75) node{latency} ;
      \draw (5.05,1.25) node{$4$} ;
      \draw[black!50] (5.05,0.75) node{$\sim$} ;
      \draw (5.05,0.25) node{$0$ --- $4$} ;

      \draw (5.7,0) -- (5.7,2.4) ;

      \draw (7.25,2.15) node{Write} ;

      \draw (6.6,1.75) node{\# messages} ;
      \draw (6.6,1.25) node{$\mathcal{O}(n)$} ;
      \draw[black!50] (6.6,0.75) node{$\sim$} ;
      \draw (6.6,0.25) node{$\mathcal{O}(n^2)$} ;

      \draw (7.5,0) -- (7.5,1.5) ;

      \draw (8.15,1.75) node{latency} ;
      \draw (8.15,1.25) node{$2$} ;
      \draw[black!50] (8.15,0.75) node{$\sim$} ;
      \draw (8.15,0.25) node{$0$} ;

      \draw (8.8,0) -- (8.8,2.4) ;
      \draw (8.9,0) -- (8.9,2.4) ;

      \draw (11,2.15) node{Snapshot} ;

      \draw (9.95,1.75) node{\# messages} ;
      \draw[black!50] (9.95,1.25) node{$\sim$} ;
      \draw (9.95,0.75) node{$\mathcal{O}\left(n^2\log n\right)$} ;
      \draw (9.95,0.25) node{$0$} ;

      \draw (11,0) -- (11,1.5) ;

      \draw (12.05,1.75) node{latency} ;
      \draw[black!50] (12.05,1.25) node{$\sim$} ;
      \draw (12.05,0.75) node{$\mathcal{O}\left(n\log(n) \right)$} ;
      \draw (12.05,0.25) node{$0$ --- $4$} ;

      \draw (13.1,0) -- (13.1,2.4) ;

      \draw (15.2,2.15) node{Update} ;

      \draw (14.15,1.75) node{\# messages} ;
      \draw[black!50] (14.15,1.25) node{$\sim$} ;
      \draw (14.15,0.75) node{$\mathcal{O}\left(n^2\log n\right)$} ;
      \draw (14.15,0.25) node{$\mathcal{O}(n^2)$} ;

      \draw (15.2,0) -- (15.2,1.5) ;

      \draw (16.25,1.75) node{latency} ;
      \draw[black!50] (16.25,1.25) node{$\sim$} ;
      \draw (16.25,0.75) node{$\mathcal{O}\left(n\log(n)\right)$} ;
      \draw (16.25,0.25) node{$0$} ;

      \draw (17.3,0) -- (17.3,2.4) ;

    \end{tikzpicture}
    }
    \caption{Complexity of several algorithms to implement a shared memory}
    \label{fig:complexity}
\end{figure}

Algorithm~\ref{algo:SCS}, like~\cite{attiya1995sharing}, uses unbounded integer values to timestamp messages. 
Therefore, the complexity of an operation depends on the number $m$ of operations executed before it, in the linear extension. 
All messages sent by Algorithm~\ref{algo:SCS} have a size of 
$\mathcal{O}\left(log(n m)\right)$. The same complexity is necessary to implement $n$ instances of a register with ABD.

In terms of local memory, due to asynchrony, in some cases $\AlgoG_i$
may contain an entry $\Algog$
for each value previously written. In that case, the space occupied by
$\AlgoG_i$ may grow up to $\mathcal{O}(m n\log m)$.
Remark though that, by Lemma~\ref{lemma:safety}, an entry $\Algog$ is
eventually removed from $\AlgoG_i$
(in a synchronous system, after $2$ time units if $\Algog.\AlgoGK = i$
or $1$ time unit if $\Algog.\AlgoGK \neq i$).
Thus, this maximal bound is unlikely to happen. Also, if all processes
stop writing (e.g. in the round based model
we discussed in Section~\ref{sec:round}), eventually $\AlgoG_i =
\emptyset$ and the space occupied by the algorithm drops down to
$\mathcal{O}(n\log m)$, which is comparable to ABD. In comparison, the
AR implementation keeps a tree containing past values from all registers,
in each register which leads to a much higher size of messages and local
memory.

\section{Conclusion}\label{sec:conclusion}

In this paper, we investigated the advantages of focusing on sequential consistency. We show that in many applications,
 the lack of composability is not a problem. The first case concerns applications built on a layered architecture and the second
 example concerns round-based algorithms where processes
access to one different sequentially consistent object in each round.

Using sequentially consistent objects instead of their linearizable counterpart can be very profitable in terms of execution time of operations.
Whereas waiting is necessary for all operations when implementing linearizable memory,
we presented an algorithm in which waiting is only required for read operations when they follow directly a write operation.
This extends the result of Attiya and Welch to asynchronous systems with crashes. Moreover, the proposed algorithm implements
a sequentially consistent snapshot memory for the same cost.

Exhibiting such an algorithm is not an easy task for two reasons. First, as write operations are wait-free, a process may
write before its previous write has been acknowledged by other processes, which leads to ``concurrent'' write operations by the same process. 
Second, proving that an implementation is sequentially consistent is more difficult than proving it is linearizable since 
the condition on real time that must be respected by linearizability highly reduces the number of linear extensions that need to be considered.

\paragraph{Acknowledgments.}
This work has been partially supported by the Franco-German ANR project DISCMAT under grant agreement ANR-14-CE35-0010-01. 

\bibliographystyle{splncs}
\bibliography{biblio}

\end{document}